\documentclass{emulateapj}		

\usepackage{natbib}				
\usepackage[caption=false]{subfig}   	
\usepackage{textcomp}

\shorttitle{AO Imaging \& Polarimetry of Hypergiants}

\begin{document}

\title{Probing Hypergiant Mass Loss with Adaptive Optics Imaging \& Polarimetry in the Infrared:  MMT-Pol\footnotemark[*] and LMIRCam observations of IRC +10420 \& VY Canis Majoris}
\footnotetext[*]{Observations reported here were obtained at the MMT Observatory, a joint facility of the Smithsonian Institution and the University of Arizona.}

\author{Dinesh P. Shenoy\altaffilmark{1}, Terry J. Jones\altaffilmark{1}, Chris Packham\altaffilmark{2}, Enrique Lopez-Rodriguez\altaffilmark{2}}
\affil{$^{1}$ Minnesota Institute for Astrophysics, University of Minnesota, 116 Church St. SE, Minneapolis, MN 55455, USA \\ email: shenoy@astro.umn.edu }
\affil{$^{2}$ Department of Physics \& Astronomy, University of Texas - San Antonio, One UTSA Circle, San Antonio TX 78249, USA}
\slugcomment{submitted to the Astronomical Journal on 2015 March 10; accepted 2015 May 16}

\begin{abstract}
We present 2 $-$ 5 $\micron$ adaptive optics (AO) imaging and polarimetry of the famous hypergiant stars IRC +10420 and VY Canis Majoris.  The imaging polarimetry of IRC +10420 with MMT-Pol at $2.2~\micron$ resolves nebular emission with intrinsic polarization of 30\%, with a high surface brightness indicating optically thick scattering.  The relatively uniform distribution of this polarized emission both radially and azimuthally around the star confirms previous studies that place the scattering dust largely in the plane of the sky. Using constraints on scattered light consistent with the polarimetry at $2.2~\micron$, extrapolation to wavelengths in the $3-5~\micron$ band predicts a scattered light component significantly below the nebular flux that is observed in our LBT/LMIRCam $3-5~\micron$ AO imaging.  Under the assumption this excess emission is thermal, we find a color temperature of $\sim$ 500 K is required, well in excess of the emissivity-modified equilibrium temperature for typical astrophysical dust.  The nebular features of VY CMa are found to be highly polarized (up to 60\%) at 1.3 $\micron$, again with optically thick scattering required to reproduce the observed surface brightness.  This star's peculiar nebular feature dubbed the ``Southwest Clump'' is clearly detected in the 3.1 $\micron$ polarimetry as well, which, unlike IRC+10420, is consistent with scattered light alone.   The high intrinsic polarizations of both hypergiants' nebulae are compatible with optically thick scattering for typical dust around evolved dusty stars, where the depolarizing effect of multiple scatters is mitigated by the grains' low albedos.
\end{abstract}


\section{INTRODUCTION}
Hypergiant stars are located near the upper limit of luminosity in the HR Diagram, with typical luminosities of $L_{\star} \sim 5 \times 10^{5}$ $L_{\sun}$ and effective photosphere radii of several AUs.  They exhibit very high mass loss rates of as much as $10^{-4}$ $M_{\sun}$ yr$^{-1}$ \citep{Danchi:1994} with some experiencing even greater discrete eruptions.  Here we present new adaptive optics (AO) infrared observations of two famous hypergiant stars IRC +10420 and VY Canis Majoris.  

IRC +10420 is a yellow hypergiant, in the rare short-lived evolutionary stage experienced by massive stars transiting to or from the red supergiant stage.  \emph{Hubble Space Telescope (HST)} visual imaging revealed a very complex circumstellar nebula with numerous condensations arrayed in jet-like structures, rays and arcs \citep{Humphreys:1997}.  $HST$ long-slit spectroscopy \citep{Humphreys:2002} combined with second epoch $HST$ imaging found that we view the star nearly pole-on, looking down on its equatorial plane \citep{Tiffany:2010}.  Recent VLTI observations of neutral and ionized gas around IRC +10420 have confirmed this geometry \citep{Oudmaijer:2013}.

VY CMa is a cool, red hypergiant.  Multi-epoch \emph{HST} imaging and polarimetry found extensive episodic mass loss with no preferred axis of symmetry \citep{Humphreys:2007, Jones:2007}.  Ground-based 2 - 5 $\micron$ AO imaging with LMIRCam \citep{Skrutskie:2010} on the Large Binocular Telescope (LBT)\footnote{\footnotesize{The LBT is an international collaboration among institutions in the United States, Italy and Germany. LBT Corporation partners are: The University of Arizona on behalf of the Arizona university system; Istituto Nazionale di Astrofisica, Italy; LBT Beteiligungsgesellschaft, Germany, representing the Max-Planck Society, the Astrophysical Institute Potsdam, and Heidelberg University; The Ohio State University, and The Research Corporation, on behalf of The University of Notre Dame, University of Minnesota and University of Virginia.}} previously reported by \citet{Shenoy:2013} found VY CMa's peculiar ``Southwest Clump'' to be a particularly dense, optically thick condensation of the ejecta.  Most recently, ALMA sub-millimeter observations have found further discrete, dense ejecta in the very close environment of the star \citep{Richards:2014, OGorman:2015}.  The mechanisms for such discrete mass loss are not understood.  Magnetic field activity analogous to coronal mass ejections may be responsible \citep{Humphreys:2007}.  Although recent $XMM-Newton$ X-ray observations of VY CMa placed constrains on the magnetic field strength at the star's surface, the star may have simply been in state of lower magnetic activity than during previous mass loss events \citep{Montez:2015}.

Because of the high contrast ratio between the star's light profile and the surface brightness of nebular material, separating infrared nebular emission from the star presents considerable difficulty.  Imaging polarimetry is a powerful technique for studying nebular emission in this regime.  Dust in the nebula scatters light from the star and polarizes it.  If the star's light is unpolarized or weakly polarized compared to the fractional polarization of the scattered star light, then imaging polarimetry can cleanly separate the faint polarized light from the star's dominating light.  We present new observations of these two hypergiant stars made with MMT-Pol, the 1 $-$ 5 $\micron$ imaging polarimeter custom designed for the 6.5 m MMT observatory on Mt. Hopkins \citep{Packham:2012} and with LBT/LMIRCam on Mt. Graham, AZ .  The observations are summarized in Table 1.  

\begin{deluxetable}{cccc}
\tablecaption{Summary of Observations}
\tablenum{1}
\tablehead{\colhead{Date} & \colhead{Instrument} & \colhead{Filter(s)} & \colhead{On-Source Time} \\ 
\colhead{(UT)} & \colhead{} & \colhead{($\micron$)} & \colhead{(s)} } 
\startdata
& & & \\
\multicolumn{4}{c}{\textbf{IRC +10420}} \\
& & & \\
2011 May 25 				& LMIRCam 				  & PAH1 (3.29) & 186 \\
\textquotesingle\textquotesingle & \textquotesingle\textquotesingle & PAH2 (3.40) & 220 \\
\textquotesingle\textquotesingle & \textquotesingle\textquotesingle & L$'$ (3.83)    & 154 \\
\textquotesingle\textquotesingle & \textquotesingle\textquotesingle & M (4.9)         & 107 \\
2012 Sep 25 & MMT-Pol & K$'$ (2.2) & 254 \\
&  &  &  \\
\tableline
& & & \\
\multicolumn{4}{c}{\textbf{VY CMa}} \\
& & & \\
2013 Oct 22 & MMT-Pol & J$'$ (1.3) & 36 \\
2014 Jan 16 & \textquotesingle\textquotesingle & 3.1 & 320 
\enddata
\end{deluxetable}

\section{OBSERVATIONS AND RESULTS}
\subsection{IRC +10420:  2.2 $\micron$ AO Imaging Polarimetry}
We observed IRC +10420 on 2012 Sep 26 (UT) with MMT-Pol \citep{Packham:2012} during its commissioning run.  MMT-Pol operates at the Cassegrain focus, where the AO-corrected beam from the secondary mirror enters the instrument with no prior off-axis reflections that might introduce spurious polarization.  A rectangular mask in the focal plane provides a 20$\arcsec$ $\times$ 40$\arcsec$ field of view.  The beam passes through a half-wave retarder (or half-wave plate, hereafter HWP), after which a Wollaston prism splits the light into orthogonal components.  These two components are  referred to as the ordinary and extraordinary rays, or $o$ and $e$ rays.  Rotating the HWP from 0$\degr$ to 45$\degr$ swaps the $o$ and $e$ rays.  Comparison of the two rays allows Stokes parameter $Q$ to be extracted while canceling out any differences in throughput between the two rays and/or differences in detector response where the rays are detected.  Similarly, rotating the HWP from 22.5$\degr$ to 67.5$\degr$ yields Stokes parameter $U$.  MMT-Pol's pixel scale is 0.043$\arcsec$ pix$^{-1}$. 

We imaged IRC +10420 through the K$'$ ($\lambda_{0}$ = 2.20 $\micron$, $\Delta\lambda$ = 0.12 $\micron$) filter at two dithered positions separated by 20$\arcsec$ on the detector array.  At each dither position images were taken as correlated double-sample (CDS) pairs, with the subtracted difference of each CDS pair yielding a single 1.33 s exposure.    For each HWP position angle 50 exposures were taken.  For image quality and flux calibration, a polarized standard star (HD 29333) and unpolarized standard star (HD 224467) were observed.  The FWHM of the PSF was 0.2$\arcsec$.  The flux calibration factors (ADUs to W cm$^{-2}$) computed using these two standards agree to within 5\%.  The instrumental polarization determined from the unpolarized standard star was 0.47\% $\pm$ 0.07\%.

Each 1.33 s exposure was examined individually.  On occasion when an increase in the width of the star profile indicated potential loss of AO lock, those exposures were discarded.  The remaining useable frames were mean combined to yield a single image in each HWP position for each of the two dithers.  The sky background and array structure were removed by subtracting dithered images; this subtraction removed many hot pixels as well.  Remaining hot pixels were replaced with the median of the surrounding 8 pixels.    The $o$-ray and $e$-ray images were extracted from these images.  The central point source of IRC +10420 in each exposure saturates the images out to a radius of $\lesssim$ 0.4$\arcsec$.  For each dither position, the 8 images (2 per HWP position) were aligned using the IRAF\footnote{\footnotesize{IRAF is distributed by the National Optical Astronomy Observatory, which is operated by the Association of Universities for Research in Astronomy (AURA) under cooperative agreement with the National Science Foundation.}} XREGISTER task.  The images were then smoothed with a Gaussian of FWHM $=$ 0.2$\arcsec$ (the beam size as estimated from the PSF star).

The normalized Stokes parameters $q \equiv Q/I$ and $u\equiv U/I$ were computed using the ratio method of \citet{Tinbergen:1996}:
\begin{eqnarray} 
q & = & \frac{R_{q}-1}{R_{q}+1}\mbox{ , for } R_{q}\equiv \sqrt{\frac{I^{o}_{00}/I^{e}_{00}}{I^{o}_{45}/I^{e}_{45}}} \\ 
u & = & \frac{R_{u}-1}{R_{u}+1}\mbox{ , for } R_{u}\equiv \sqrt{\frac{I^{o}_{22}/I^{e}_{22}}{I^{o}_{67}/I^{e}_{67}}}
\end{eqnarray}
where the $o$ and $e$ superscripts on each of the intensities $I$ refer to the ordinary and extraordinary rays emerging from the Wollaston prism, and the subscripts refer to the four HWP orientation angles $00.0\degr$, $45.0\degr$, $22.5\degr$, and $67.5\degr$.  Dividing the intensities to form ratios $R_{q}$ and $R_{u}$ minimizes potential differences in the Wollaston prism's relative throughput on the two rays, relative differences in the response of the array locations where the two rays are detected and variations of sky transmission during the time between when the images are taken through the paired orientations of the HWP.  The normalized Stokes parameters $q$ and $u$ computed with this ratio method are algebraically equivalent to the standard definitions: 
\begin{equation} 
q = \frac{Q}{I} \equiv \frac{I_{0} - I_{90}}{I_{0} + I_{90}} \hspace{12pt} ; \hspace{12pt} u = \frac{U}{I} \equiv \frac{I_{45} - I_{135}}{I_{45} + I_{135}}
\end{equation}
where $I_{\phi}$ are the intensities transmitted through a linear polarizer (analyzer) with its transmission axis at angles $\phi$ = 0$\degr$, 90$\degr$, 45$\degr$ and 135$\degr$, and $I$ is the total intensity.  The fractional polarization $p$ and position angle (PA) $\theta$ are then computed with:
\begin{equation} 
p = \sqrt{q^{2}+u^{2}} \hspace{12pt} ; \hspace{12pt} \theta = \frac{1}{2}\tan^{-1}\left( \frac{u}{q} \right)
\end{equation}
The polarized intensity is the product of the fractional polarization and total intensity:  $I_{P} = p \cdot I$.  An S/N cut of 4-$\sigma$ or 8-$\sigma$ with respect to background fluctuations in polarized intensity off the source was selected as the cut-off for the display of polarization vectors.  A 4-sigma cut is equivalent to Òde-biasingÓ by a factor of less than 0.03 (see Equation (A3) of \citet{Wardle:1974}), which has no impact on our science.

The position angle $\theta$ computed using Equation (4) is in array coordinates.  To convert to degrees East-of-North on the sky requires the addition of a calibration offset angle $\Delta\theta$, which is normally determined from observations of one or more polarized standard stars.  Observations of the polarized standard star HD 29333 during the commissioning run did not provide a meaningful calibration offset due to non-photometric conditions at the time the standard was observed and inadequate integration time.   For our observations of IRC +10420 the uncorrected polarization vectors' orientation is physically implausible.   Dust distributed around and illuminated by a single central source is expected to show a centro-symmetric scattering pattern, and previous observations of IRC +10420 by \citet{Kastner:1995} find such a pattern.  An offset of $\Delta\theta_{K'} = 58\degr$ when added to each of our uncorrected polarization vectors achieves a centro-symmetric pattern in all directions around the star.  We have therefore adopted this offset value.

\begin{figure*}
\centering
\includegraphics[scale=0.38]{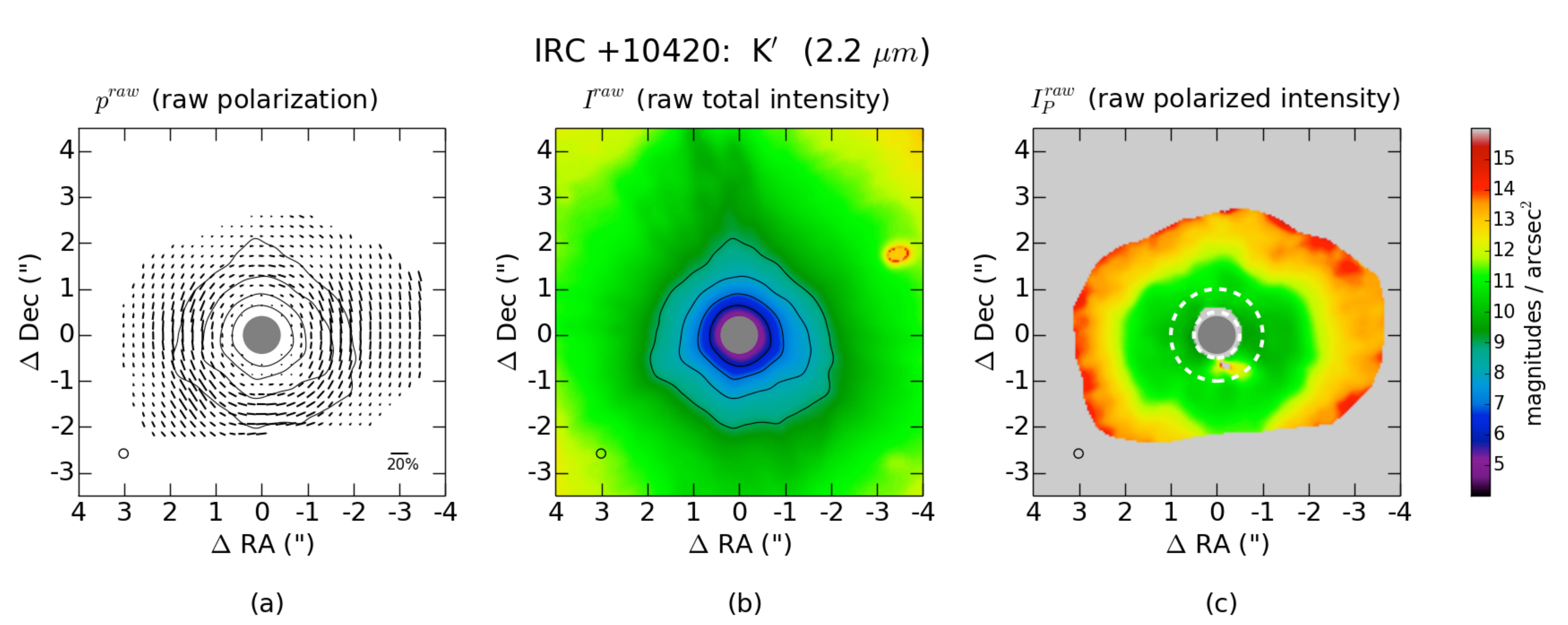}	
\caption{MMT-Pol observations of IRC +10420 at $\lambda$ = K$'$ (2.2 $\micron$).  North is up and East is left.   The center is masked out to a radius of 0.4$\arcsec$ where the images saturated.  The circle in the lower left is the beam size (0.2$\arcsec$ = FWHM of PSF). \textbf{(a):}  The vectors are the raw polarization $p^{raw}$ for a cut in polarized intensity of 4-$\sigma$ with respect to background fluctuations off the source.  The overlaid contours show the raw total intensity (total = unpolarized + polarized) from 9 to 6 mag arcsec$^{-2}$, progressing inwards in steps of $-$1 mag arcsec$^{-2}$.  A portion of the image was masked in the south where faint artifacts which appear to have been caused by internal reflections of the heavily overexposed central point source within the instrument optics cause spurious polarization.  The vectors show a centro-symmetric pattern characteristic of scattering by circumstellar dust.  The raw polarization rises to about 15\% at a radius of $\sim$ 1.7$\arcsec$ around most of the star.  A length scale for the polarization vectors is given in the lower right of the image.   \textbf{(b):}  Raw total intensity $I^{raw}$ in mag arcsec$^{-2}$, with the same overlaid contours as left image.  \textbf{(c):}  Raw polarized intensity $I^{raw}_{P}$ = $p^{raw}\cdot I^{raw}$ in mag arcsec$^{-2}$.  The dashed-line annulus indicates the region of overlap-of-coverage (OCR) with the $3-5~\micron$ LMIRCam observations (Fig. \ref{LMIRCam_ims}).   A colorbar for the surface brightnesses is on the far right.\\ \\} 
\label{IRC_pol_map}
\end{figure*}

In Figure \ref{IRC_pol_map}(a) we display our resulting $p^{raw}$, the raw polarization around IRC +10420 prior to separation into stellar and nebular components. By ``raw'' we mean uncorrected for any underlying flux from the star in the wings of the PSF. With the adopted PA offset, the raw polarization is consistent with the centro-symmetric pattern characteristic of scattering of light from the central star by circumstellar dust.  The raw percentage of polarization increases with radius, rising to a maximum of 15\% at 1.7$\arcsec$. This suggests the central star profile dilutes a higher intrinsic polarization of the nebula; this effect is corrected for in Section \S 3.1 below. In Figure \ref{IRC_pol_map}(b) we display a flux-calibrated image of the raw intensity $I^{raw}$ with the color scale in mag arcsec$^{-2}$ for comparison with the contours in 1(a).  In Fig. \ref{IRC_pol_map}(c)  we display the product of these two images, which is the polarized intensity $I^{raw}_{P}$ = $p^{raw}\cdot I^{raw}$.  While the raw intensity is dominated by the star and thus drops off smoothly with radius, the raw polarized intensity image shows a region of relatively flat, extended emission from $0.5\arcsec$ out to a radius of $\sim$ 2.5$\arcsec$.  This corresponds to emission from extended nebular material, not the stellar PSF.

\subsection{IRC +10420:  3 - 5 $\micron$ AO Imaging}
IRC +10420 was observed on 2011 May 25 UT during the commissioning of LMIRCam, the 2 $-$ 5 $\micron$ high-resolution camera on the Large Binocular Telescope (LBT) \citep{Skrutskie:2010}.  During commissioning a single 8.4 m primary mirror was used, in conjunction with the deformable AO secondary.  This system achieves near-diffraction limited imaging and high sensitivity in the thermal infrared due to minimizing the number of reflecting surfaces and keeping all mirrors after the tertiary mirror at cryogenic temperatures.  IRC +10420 was observed through two narrow-band filters:  PAH1 ($\lambda_{0}$ = 3.29 $\micron$, $\Delta\lambda$ = 0.02 $\micron$), PAH2 ($\lambda_{0}$ = 3.40 $\micron$, $\Delta\lambda$ = 0.02 $\micron$) as well as two broad-band filters:  L$'$ ($\lambda_{0}$ = 3.8 $\micron$, $\Delta\lambda$ = 0.3 $\micron$) and M ($\lambda_{0}$ = 4.9 $\micron$, $\Delta\lambda$ = 0.3 $\micron$).  LMIRCam has a field of view approximately 10$\arcsec$ $\times$ 10$\arcsec$, with a pixel scale of 0.011$\arcsec$ pixel$^{-1}$.  The images of IRC +10420 were dithered for background sky subtraction.  For the two PAH filters, the star $\eta$ Aquilae provides a PSF ($\eta$ Aql is point-like in the 3 - 8 $\micron$ range \citep{Barmby:2011}), while for the L$'$ filter the star BD +35 2435 is the PSF.  

\begin{figure} 
\centering
\includegraphics[scale=0.28]{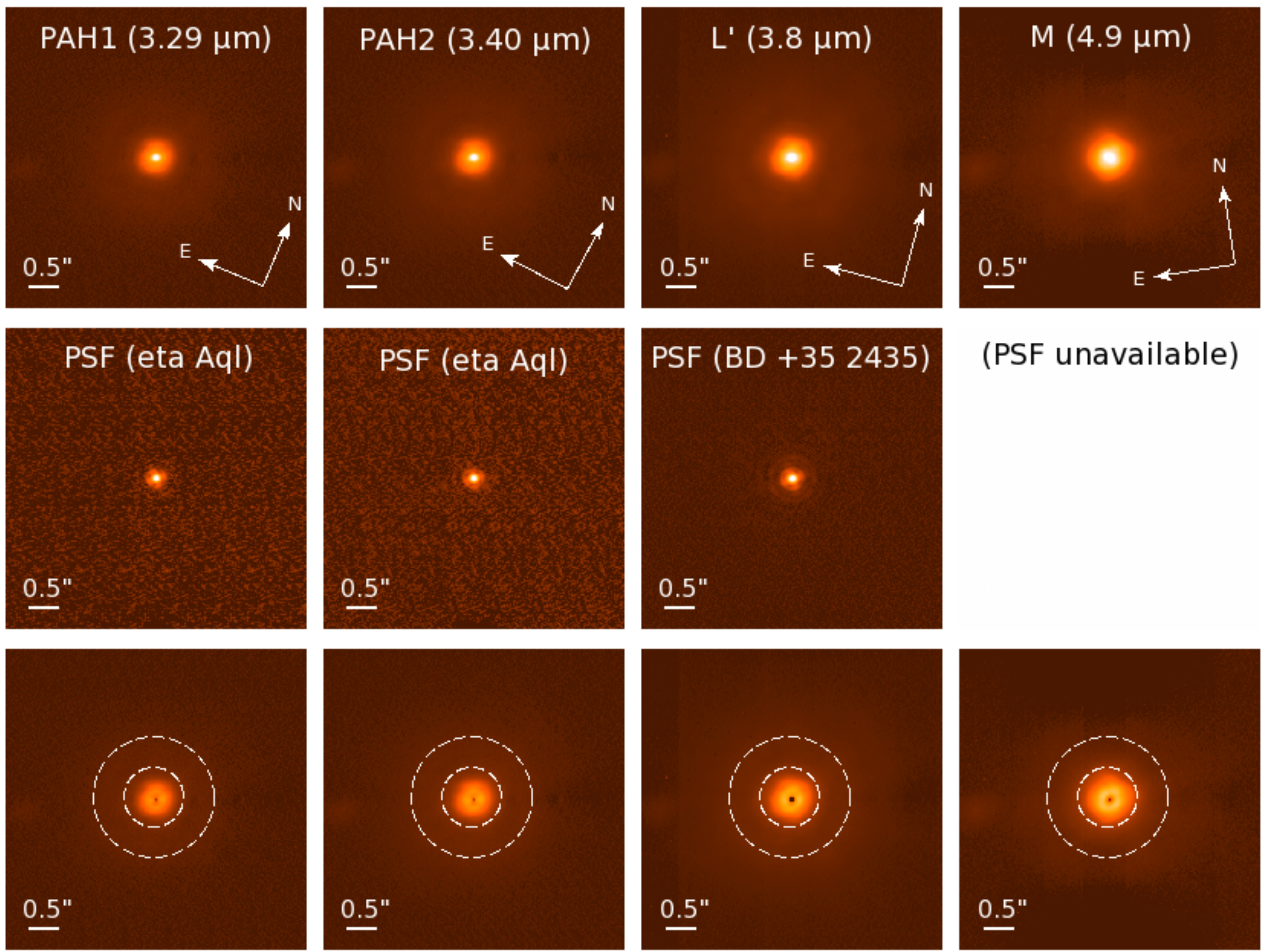}	
\caption{IRC +10420:  3 - 5 $\micron$ adaptive optics images made with LBT / LMIRCam.  \textbf{Top Row:}  IRC +10420, with the filter for each column of images as indicated.  Each FOV is 5$\arcsec$ $\times$ 5$\arcsec$, with square-root scaling to emphasize the extended nebular emission.  \textbf{Middle Row:}  PSF stars.  For filters PAH1 and PAH2 the PSF is the star $\eta$ Aql (FWHM = 0.10$\arcsec$ in each filter), while at L$'$ the PSF is BD +35 2435 (FWHM = 0.11$\arcsec$).  There is no PSF image for the M filter from that night.  \textbf{Bottom Row:}  Same as the top-row images, after subtracting a scaled Moffat fit to each filter's PSF.  To simulate a PSF for the M filter, the FWHM of the Moffat function used at L$'$ was scaled by the ratio $\lambda_{M}/\lambda_{L'}$.  The top and middle row images are each stretched to the maximum pixel value at the center.  The subtracted images of IRC +10420 in the bottom row are stretched to the same maximum value as the top row images.  The dashed-line annulus on the subtracted images is the region of overlap-of-coverage (OCR) with the 2.2 $\micron$ polarimetry.  It spans from 0.5$\arcsec$ (where the 2.2 $\micron$ polarimetry is outside the saturated stellar image) out to 1.0$\arcsec$, the 3-$\sigma$ level in the PAH1 (3.29 $\micron$) and M images.\\} 
\label{LMIRCam_ims}
\end{figure}

In Figure \ref{LMIRCam_ims} we display the four LMIRCam images of IRC +10420 along with the PSF stars for comparison.  In the bottom row of the figure, we have subtracted a scaled Moffat profile fitted to the PSF at each wavelength.  For the M filter we radially scaled the L$'$ PSF profile by the ratio of $\lambda_{M}$ to $\lambda_{L'}$.  In all four filters IRC +10420 is substantially extended compared to the PSF.  

We used two methods to determine a mean flux calibration factor in each filter.  For the first method we equated the integrated intensity in ADUs per second out to the 3-$\sigma$ level in each image (i.e., entire source = star + extended emission) with the specific flux $F_{\nu}$ from the ISO SWS spectrum for IRC +10420.  For the second method, we calibrated the ADUs per second against those of the observed standard stars.  For the flux of the standard star $\eta$ Aql at the PAH1 \& PAH2 filter wavelengths we interpolated between its reported magnitudes in the 2MASS Ks filter and IRAC Band 1 \citep{Marengo:2010}.  For the flux of the standard star BD +35 2435 at L$'$ we interpolated between its WISE W1 and W2 magnitudes.  In the absence of a standard star observation in the M filter, we repeated the first method with comparison to IRC +10420's magnitude at M from the large, single aperture photometry of \citet{Jones:1993}.  For each filter we use the mean calibration factors from the two methods, with estimated uncertainties of 15\%, 10\%, 10\% and 5\% in the PAH1, PAH2, L$'$ and M filters respectively. 

\subsection{VY CMa:  J$'$ (1.3 $\micron$) and 3.1 $\micron$ Imaging Polarimetry}
\begin{figure*}
\centering
\includegraphics[scale=0.35]{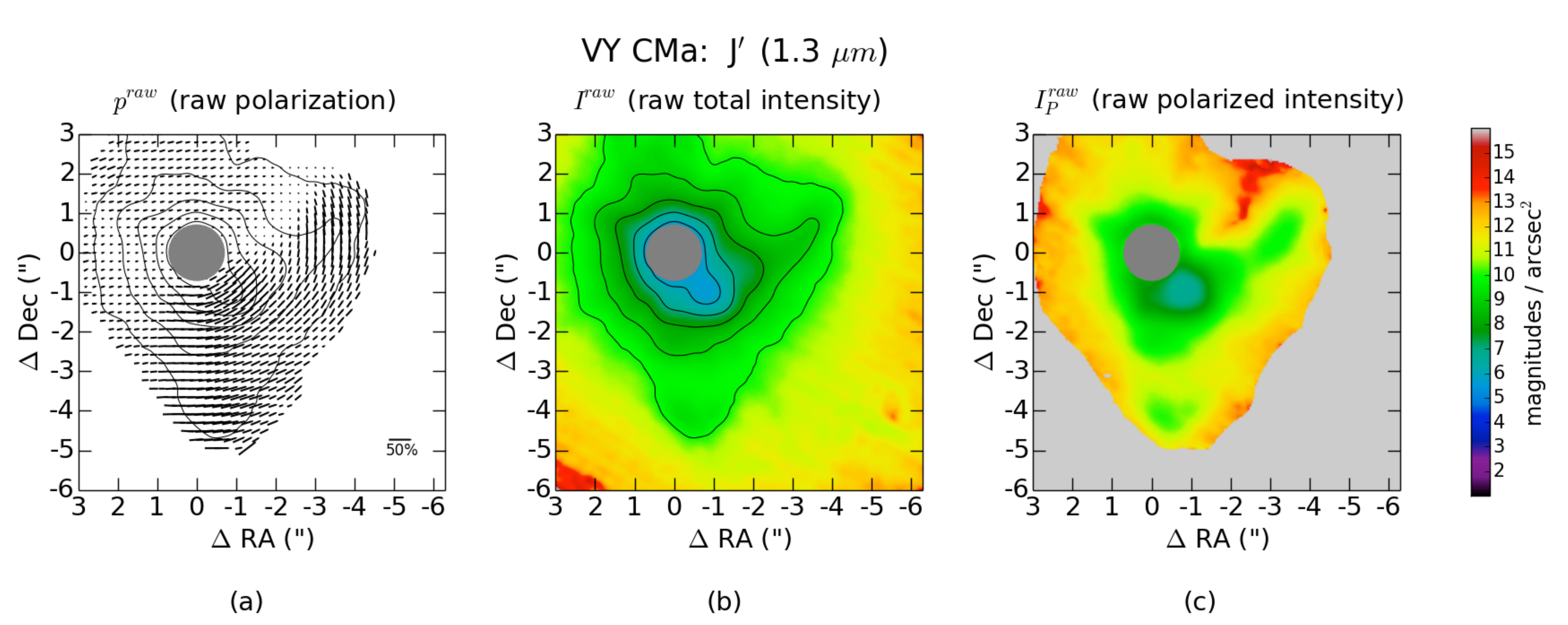}
\caption{MMT-Pol observations of VY CMa at $\lambda$ = J$'$ (1.3 $\micron$).  North is up and East is left.   The center is masked out to a radius of 0.6$\arcsec$ where the star saturates.   \textbf{(a):}  The vectors are the raw polarization $p^{raw}$ for a cut in polarized intensity of 4-$\sigma$ with respect to background fluctuations off the source.  The overlaid contours show the raw total intensity (total = unpolarized + polarized).   The outermost contour is 10 mag arcsec$^{-2}$, with contours progressing inwards in steps of $-$1 mag arcsec$^{-2}$.  The extended shape of the outermost contour towards the North is due to a faint artifact which appears to have been caused by internal reflections of the heavily overexposed central point source within the instrument optics.  In contrast the extension to the South is real, coinciding with the location of the distinct component of the ejecta previously identified as Arc 2 by \citet{Humphreys:2007} (see Fig. \ref{VY_HST_comparison}(a) below). The polarization vectors show a centro-symmetric pattern characteristic of scattering by circumstellar dust.  \textbf{(b):}  Raw total intensity $I^{raw}$ in magnitudes arcsec$^{-2}$, with the same overlaid contours as left image.  \textbf{(c):}  Raw polarized intensity $I^{raw}_{P}$ = $p^{raw}\cdot I^{raw}$ in mag arcsec$^{-2}$.  A colorbar for the surface brightnesses is on the far right.} 
\label{VY_pol_map_J}
\end{figure*}

\begin{figure*}	
\centering
\subfloat[][]{\label{VY_HST_F1042M} \includegraphics[scale=0.23]{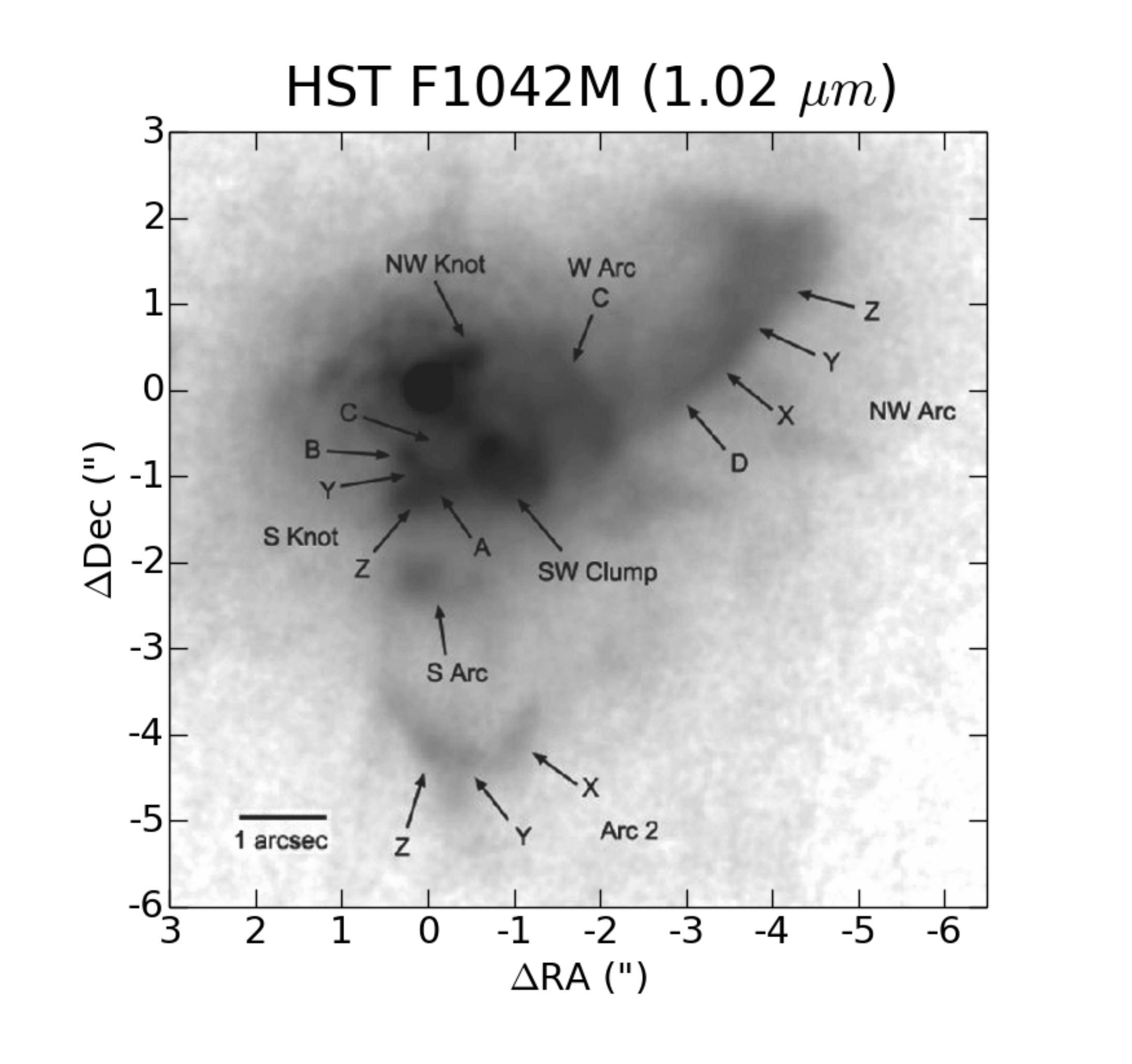} }
\subfloat[][]{\label{VY_pol_map_HST} \includegraphics[scale=0.23]{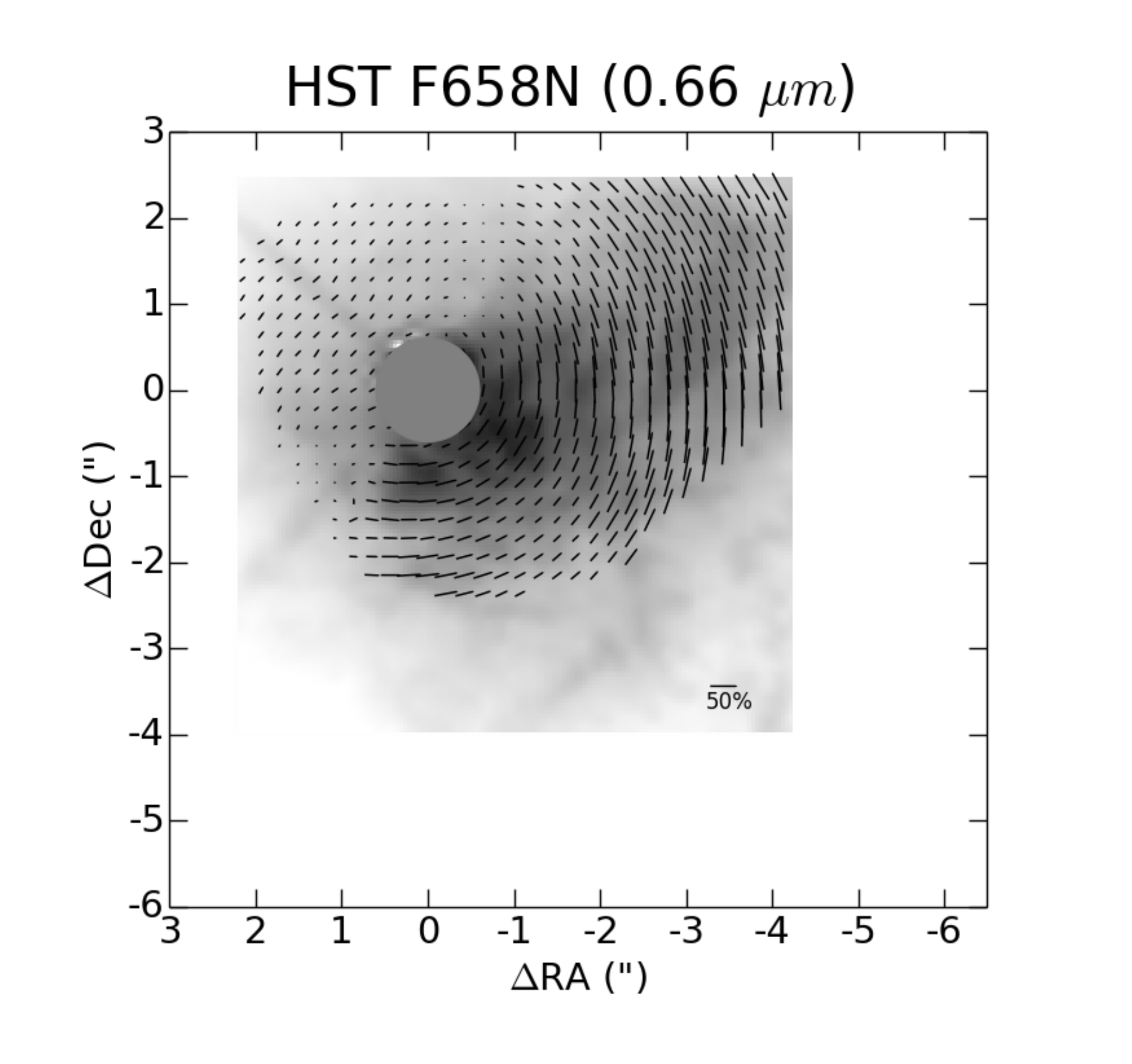} }
\subfloat[][]{\label{VY_J_solo} \includegraphics[scale=0.23]{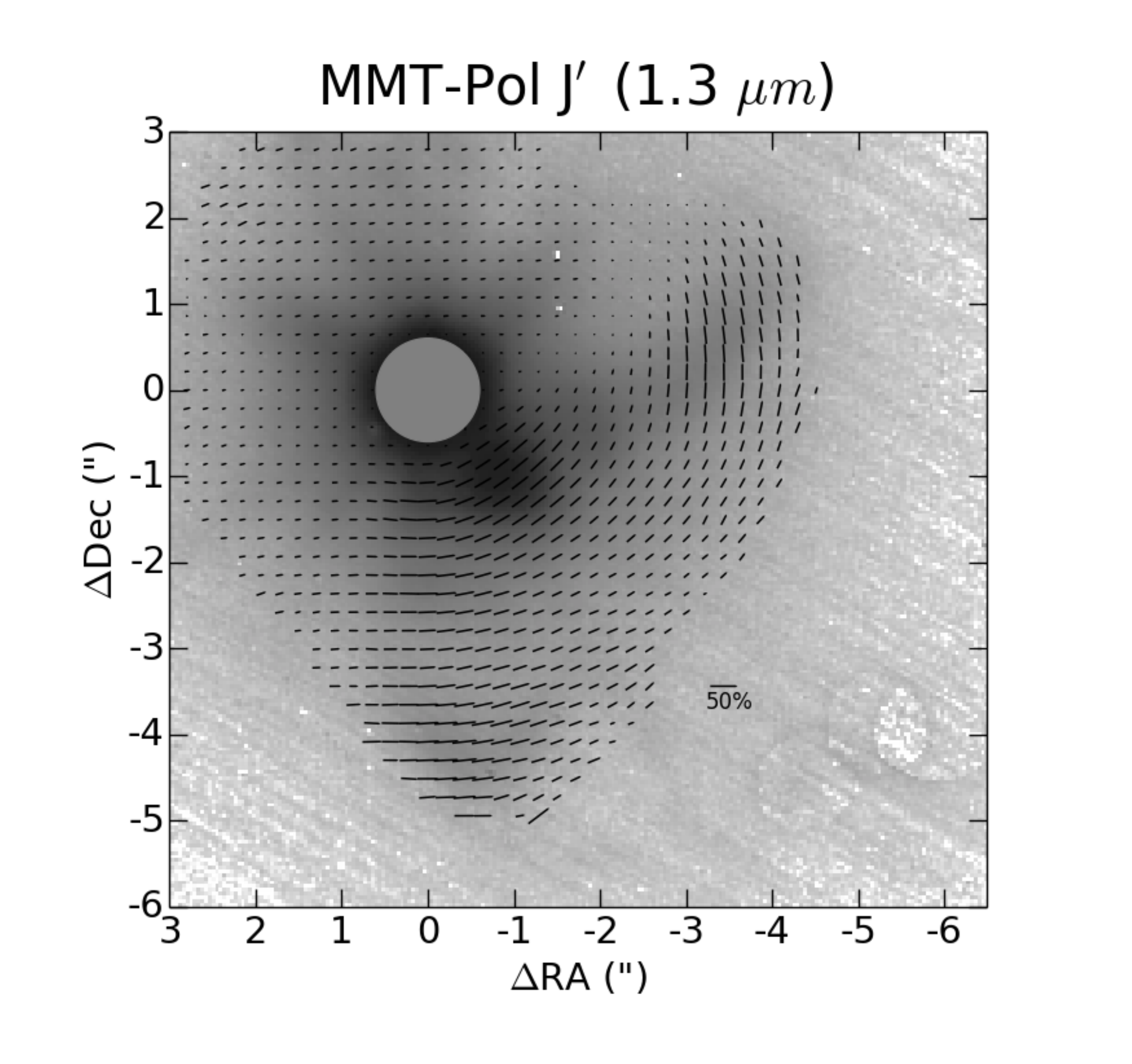} }
\caption{VY CMa:  \textbf{(a)}  \emph{HST} F1042M (1 $\micron$) image reproduced from \citet{Humphreys:2007} identifying the NW Arc, Arc 2, S Knot, S Arc and SW Clump. \textbf{(b)} \emph{HST} visual (0.66 $\micron$) polarimetry replotted using data from \citet{Jones:2007}.  \textbf{(c)} MMT-Pol J$'$ (1.3 $\micron$) polarimetry (this work).\\} 
\label{VY_HST_comparison}
\end{figure*}

We observed VY CMa with MMT-Pol on 2013 Oct 22 UT through a narrow-band J$'$ ($\lambda_{0}$ = 1.3 $\micron$, $\Delta\lambda$ = 0.1 $\micron$).  Six sets of dithered images were taken with a 1 s exposure time per HWP position.   Data reduction was performed  as discussed above for the MMT-Pol images of IRC +10420.  The unpolarized standard star HD 224467 provided the PSF and flux-calibration.  The offset $\Delta \theta_{J'}$ added to all polarization vectors was chosen to yield a centro-symmetric scattering pattern consistent with the \emph{HST} visual polarimetry in \citet{Jones:2007}.  We estimate our photometric uncertainty in the flux-calibration from the two standards at 7\%.  We display our J$'$ polarimetry in Figure \ref{VY_pol_map_J}, and a comparison to the $HST$ visual polarimetry in Figure \ref{VY_HST_comparison}.

\begin{figure*}	
\centering
\includegraphics[scale=0.35]{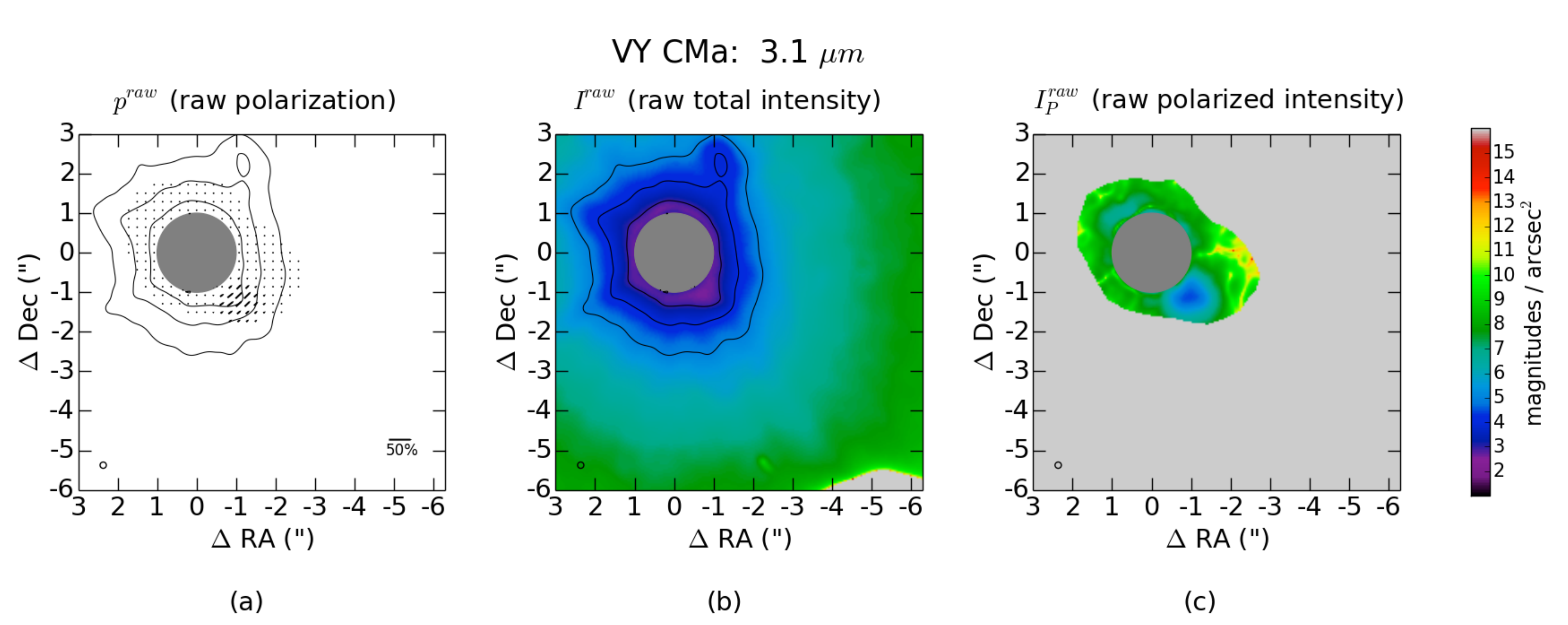}
\caption{MMT-Pol observations of VY CMa at $\lambda$ = 3.1 $\micron$.  North is up and East is left.   The center is masked out to a radius of 1.0$\arcsec$ where the star saturates.   \textbf{(a):} The vectors are the raw polarization $p^{raw}$ for a cut in polarized intensity of 8-$\sigma$ with respect to background fluctuations off the source.  The overlaid contours show the raw total intensity (total = unpolarized + polarized).  The outermost contour is 5 mag arcsec$^{-2}$, with contours progressing inwards in steps of $-$1 mag arcsec$^{-2}$.  \textbf{(b):}  Raw total intensity $I^{raw}$ in magnitudes arcsec$^{-2}$, with the same overlaid contours as left image.  \textbf{(c):}  Raw polarized intensity $I^{raw}_{P}$ = $p^{raw}\cdot I^{raw}$ in mag arcsec$^{-2}$.  A colorbar for the surface brightnesses is on the far right.  The raw polarization at the location of the SW Clump is 20\%.\\} 
\label{VY_pol_map_3p1}
\end{figure*}

We subsequently observed VY CMa through MMT-Pol's narrow-band 3.1 $\micron$ filter ($\Delta\lambda$ = 0.1 $\micron$) on 2014 Jan 16 UT. The PSF and flux-calibration were obtained from observations of HD 104624 and HD 29333.  We estimate our photometric uncertainty in the flux-calibration from these two standards at 5\%.  For the 3 - 5 $\micron$ range, MMT-Pol uses a retarder which provides a retardance of 180$\degr$ (half-wave) at 3.6 $\micron$.  For 3.1 $\micron$ light, the waveplate's retardance is 209$\degr$.   Defining efficiency as the ratio of the input polarized intensity to the measured polarized intensity, the difference in retardance lowers the efficiency of the polarization measurement at 3.1 $\micron$ by a factor of 0.94.  We have corrected the raw polarized intensity by this 6\% factor during the reduction of the 3.1 $\micron$ images.  The data reduction is otherwise the same as for IRC +10420, except we applied a S/N cut-off of 8-$\sigma$ in polarized intensity to mitigate faint artifacts from the heavily over-exposed central star.   A polarization vector offset has been added which achieves a centro-symmetric pattern.  We display our result in Figure \ref{VY_pol_map_3p1}. 

\section{ANALYSIS \& DISCUSSION}
\subsection{IRC +10420:  Nebula's Intrinsic Polarization at K$'$}
In their large-aperture polarimetry of IRC +10420 \citet{Jones:1993} observed a fractional polarization of 1\% at K$'$ for the combined star plus any unresolved nebula.  The sub-arcsecond resolution now possible with MMT-Pol allows us to resolve nebular emission for radii greater than about 0.4$\arcsec$ (2000 AU), limited mostly by the exceedingly bright central star, which necessarily saturates in the integration times required to bring up the nebulosity.  In our image the raw fractional polarization $p^{raw}$ varies with radius, rising with increasing distance from the star up to a peak of about 15\% at 1.7$\arcsec$ and then dropping off.  This behavior indicates that flux in the PSF from the weakly polarized central star is diluting a higher intrinsic polarization of the nebula nearer the star.  

\begin{figure*}  
\centering
\includegraphics[scale=0.35]{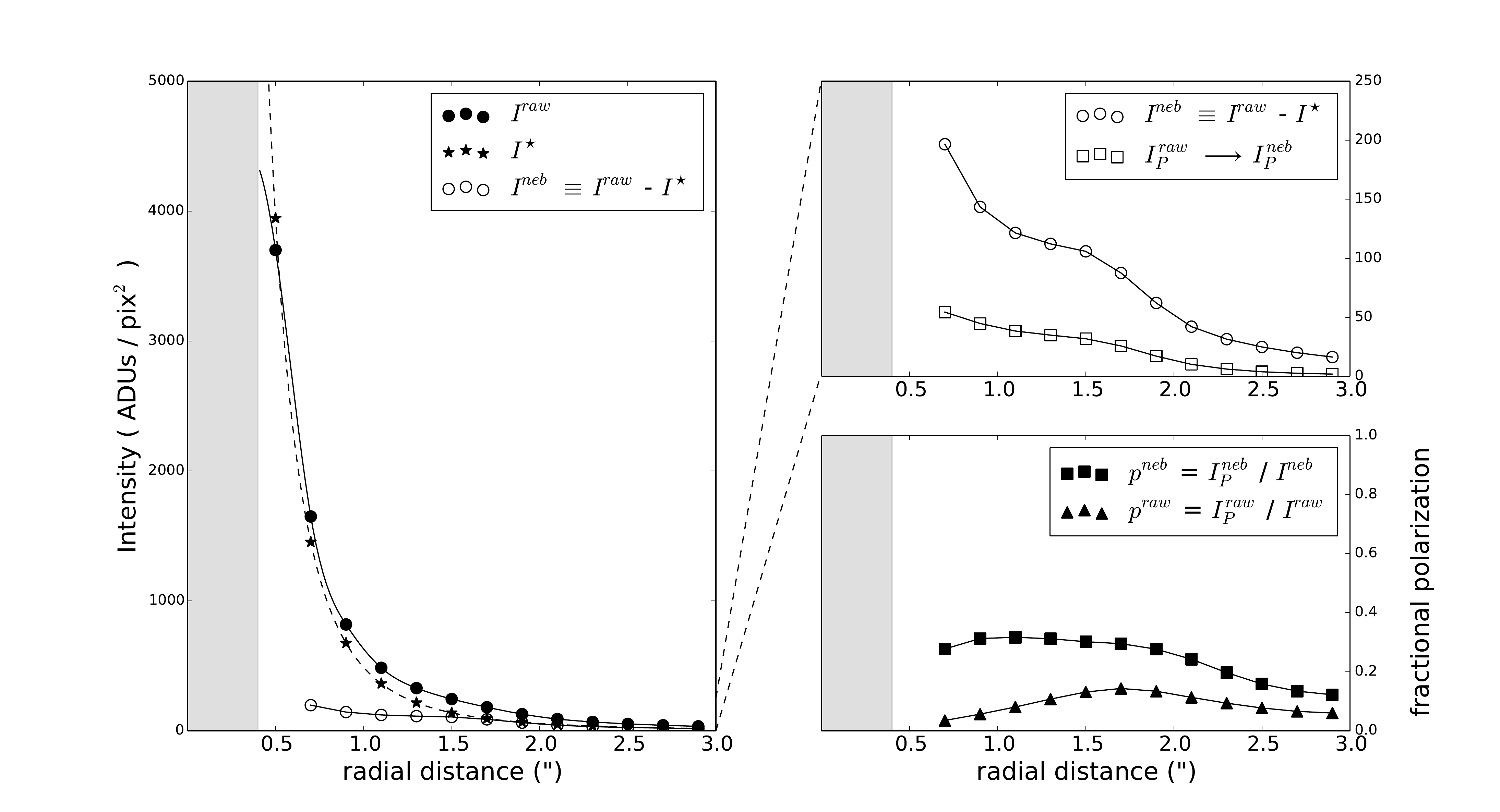}	
\caption{Estimating the intrinsic polarization of IRC +10420's nebula at K$'$ (2.2 $\micron$).  \textbf{Left:}  The filled circles are the azimuthal average of the raw intensity $I^{raw}$ to the East and West of IRC +10420, using radial bins equal to the beam size (0.2$\arcsec$).  The dashed line is a Moffat profile fitted to the PSF, scaled to represent the star's profile $I^{\star}$.  The difference (open circles) is deemed to be the nebula's average profile $I^{neb}$.  \textbf{Upper Right:}  A zoomed view of the nebula's average profile $I^{neb}$ (open circles).  The open squares are the average profile of the raw polarized intensity $I^{raw}_{P}$ (see Figure \ref{IRC_pol_map}(c)), which is then assumed to be solely nebular emission:  $I^{raw}_{P}$ $\longrightarrow$ $I^{neb}_{P}$.  \textbf{Lower Right:}  The nebula-only polarized intensity is then divided by the calculated nebula's total intensity to yield the nebula's intrinsic polarization $p^{neb}$ = $I^{neb}_{P}$ / $I^{neb}$ (filled squares), compared to the raw polarization $p^{raw}$ (triangles).  This removal of the star's light profile shows that on average the nebula's intrinsic polarization is at least a factor of 2 higher than the raw polarization and relatively constant from 0.5$\arcsec$ $-$ 2$\arcsec$. \\} 
\label{az_avg_plot}
\end{figure*}

To assess the nebula's intrinsic polarization, we first estimate the nebula's total (unpolarized + polarized) intensity $I^{neb}$  by subtracting a representative star profile from the raw intensity $I^{raw}$.  On the left-hand side of Figure \ref{az_avg_plot} we plot the azimuthal average profile of $I^{raw}$ for the regions to the East and West of the star, excluding the regions to the North and South where faint artifacts from the star are present.  The star's light profile  $I^{\star}$ is made using a Moffat function fitted to the PSF so that $I^{\star}$ can be extended out into the nebula.  The fitted star profile has been scaled vertically so that it does not over-subtract by a radius of about 0.5$\arcsec$.  This scaling  consistent with the difference in observed K magnitudes of the PSF star and IRC +10420, assuming K = 3.5 for the latter \citep{Oudmaijer:2009}.  The difference of the observed and star profiles is deemed to be the nebula's total intensity:  $I^{neb}$ $\equiv$ $I^{raw}$ $-$ $I^{\star}$, which is depicted in Figure \ref{az_avg_plot} with open circles.

For the nebula's polarized intensity $I^{neb}_{P}$ we assume that all of the raw polarized emission in Figure \ref{IRC_pol_map}(c) is entirely from the nebula:  $I^{\star}_{P}$ $\ll$ $I^{neb}_{P}$ so that $I^{raw}_{P}$ $\longrightarrow$ $I^{neb}_{P}$.  This only slightly overestimates $I^{neb}_{P}$ within a radius of about 1$\arcsec$ and becomes increasingly accurate as radius increases.  For example with a 1\% polarized star whose profile follows the PSF profile, by a radius of 1$\arcsec$ the star's polarized intensity $I^{\star}_{P}$ is $\sim$ 10\% of $I^{raw}_{P}$ and falls off rapidly as the star's light profile descends going farther into the nebula region.  The radial profile of $I^{neb}_{P}$ is plotted with open squares in the upper right of Figure \ref{az_avg_plot} along with the nebula's total intensity $I^{neb}$.  The ratio of these intensities is used to compute the nebula's intrinsic fractional polarization, which by definition is $p^{neb}$ $\equiv$ $I^{neb}_{P}$ / $I^{neb}$.  This intrinsic polarization $p^{neb}$ is plotted with solid black squares in the lower right of Figure \ref{az_avg_plot}.  This removal of the star's light profile indicates that the nebula's intrinsic polarization is about 30\% over a broad radial range from 0.5$\arcsec$ to $\sim$ 2$\arcsec$.

Based on radial velocity and proper motion measurements, \citet{Tiffany:2010} found that the ejecta seen in $HST$ visual images is moving mostly in the plane of the sky, and concluded that IRC+10420 is nearly pole-on and we are looking down onto an equatorial dust distribution. The relatively high fractional polarization (30\%) at K$'$ that we find for the nebulosity surrounding the star is consistent with this model. For pure Rayleigh singly scattered light (size parameter $2\pi a/\lambda \ll 1$), polarization as high as 100\% could in principle be expected for optically thin dust located exactly in the plane of the sky with respect to the star (scattering angle $\Theta$ = 90$\degr$).  Realistically the maximum possible polarization is likely lower for one or more reasons.  For a distribution of grain sizes that includes larger grains with $2\pi a/\lambda$ on the order of 1, Mie theory applied to spherical grains indicates the maximum polarization decreases below 100\%.  Consider for  example a Gaussian distribution of grain sizes with mean radius $\bar{a}$ = 0.3 $\micron$ and standard deviation $\sigma_{a}$ = 0.15 $\micron$, with astronomical silicate optical constants at 2.2 $\micron$ of $n$ = 1.72, $k$ = 0.035 \citep{Draine:1984}.  Computing the maximum polarization for such a distribution with the BHMIE subroutine \citep{BohrenHuffman:1983} yields a maximum polarization of $\sim$ 60$\degr$, occurring at a scattering angle slightly behind the plane of the sky.  Alternately, for a distribution of scattering angles about 90$\degr$ (a flared disk, for example) and multiple scatters, one would expect a lower fractional polarization than the ideal maximum. Since the nebulosity as seen in polarized intensity is distributed in a nearly circular pattern around the star and has a near constant fractional polarization, it can not be in a disk or torus that is tilted to the line of sight. For such a disk or torus the projected shape would be elliptical and the scattering angles would vary with azimuthal angle. The only other possible geometry that could explain the observed morphology and polarization of the nebulosity surrounding IRC+10420 is a thin cone opening towards (or away) from us at a uniform scattering angle and surface brightness with distance from the star.  We find this to be implausible in light of the radial velocity measurements, although not ruled out by our polarimetry.

\subsection{VY CMa: Intrinsic Polarization of Nebular Features at J$'$ (1.3 $\micron$) and 3.1 $\micron$}
The polarized intensity images of VY CMa in Figures \ref{VY_pol_map_J}(c) \& \ref{VY_pol_map_3p1}(c) reveal several distinct features of its complex nebula.  Using the names designated by \citet{Humphreys:2007} in their $HST$ F1042M (1 $\micron$) image (reproduced in Figure \ref{VY_HST_F1042M}), at J$'$ we observe the Northwest Arc, Arc 2, the Southwest Clump, the South Knot, and the South Arc.   The first two features are located sufficiently distant from the star that their raw polarization is taken to be their intrinsic polarization.  The raw polarimetry at J$'$ (1.3 $\micron$) shows the NW Arc is $\sim$35\% polarized, while Arc 2's polarization rises as high as 45\%.  These and subsequently discussed values are summarized in Table 2.

\begin{figure*}
\centering
\subfloat[][] {\label{VY_pol_vecs_3p1} \includegraphics[scale=0.35]{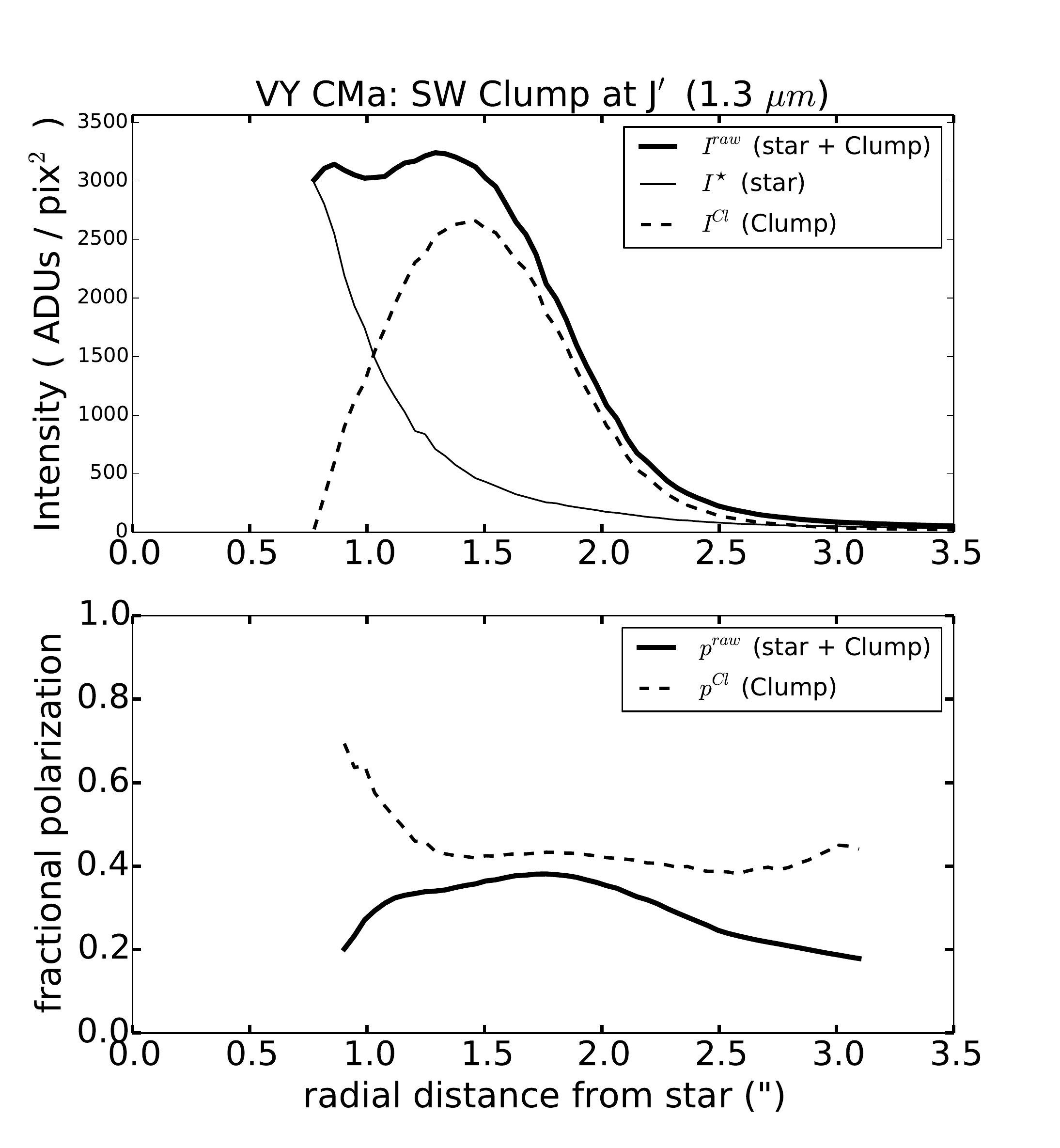}} 
\subfloat[][] {\label{VY_temp_plot_3p1} \includegraphics[scale=0.35]{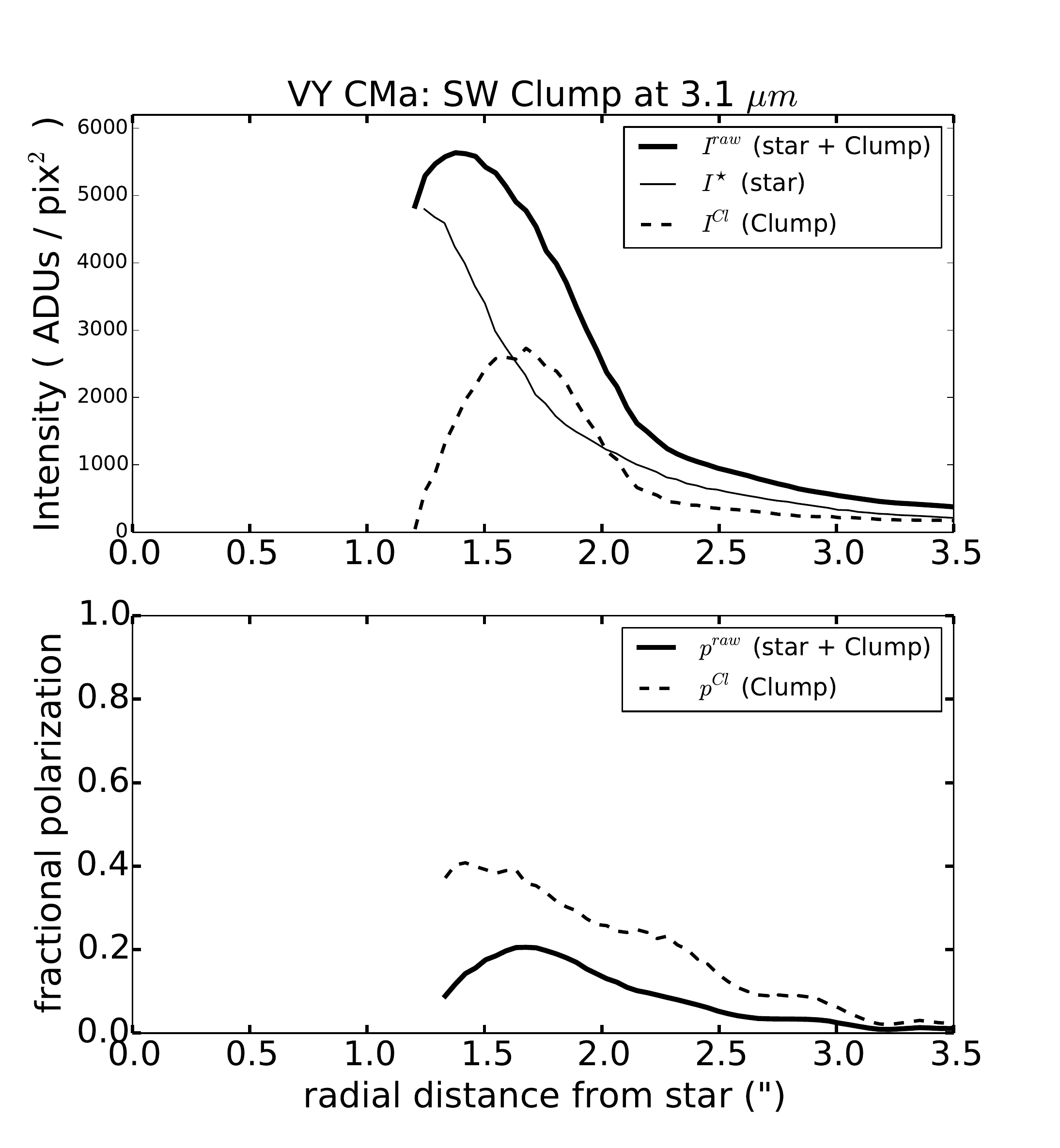}}
\caption{VY CMa:  Photometric cuts through the SW Clump at J$'$ (1.3 $\micron$) (left column) and 3.1 $\micron$ (right column).  The SW Clump's proximity to the star results in the star's light profile diluting the Clump's intrinsic polarization.  \textbf{Top row of (a) \& (b)}:  For each  indicated filter the thick solid line is the raw (total) intensity along a cut through the Clump at a position angle of $-135\degr$ E of N.  The thin solid line is a cut at a position angle of +148$\degr$ E of N, which is taken to represent the star's light profile into the region of the Clump.  The dashed line is the difference of the two.  \textbf{Bottom row of (a) \& (b)}:  For each filter the thick solid line is the raw fractional polarization $p^{raw}$ along the same cut through the location of the Clump.  The Clump's intrinsic fractional polarization $p^{Cl}$ is estimated from scaling up $p^{raw}$ by the ratio of $I^{raw}$ / $I^{Cl}$ from the plot above.} 
\label{SW_Clump_cuts}
\end{figure*}

For the SW Clump, S Knot and S Arc, the intrinsic polarization of each is diluted by the star to varying extents.  Applying the same procedure as in \S 3.1, we assume the raw polarized intensity is solely from the nebular feature so that $I^{raw}_{\lambda,P}$ $\longrightarrow$ $I^{neb}_{\lambda,P}$.  We estimate the feature's total intensity $I^{neb}_{\lambda}$ by subtracting from the raw intensity the profile of the star's light, which we represent using a photometric cut at position angle $+148\degr$ East of North.  Each nebular feature's intrinsic polarization is the ratio:  $p^{neb}_{\lambda}$ $\equiv$ $I^{neb}_{\lambda,P}$ /  $I^{neb}_{\lambda}$.  For the SW Clump as shown in Figure \ref{SW_Clump_cuts}(a) for example,  this raises its J$'$ raw fractional polarization slightly to an intrinsic polarization of 40\%.    For the S Knot and S Arc the same procedure raises the raw polarization of each from 40\% to an intrinsic polarization of as much as 60\%.  

At 3.1 $\micron$ the SW Clump is the sole feature observed in polarized intensity (Figure \ref{VY_pol_map_3p1}(c)).  Per Figure \ref{SW_Clump_cuts}(b) subtracting the star profile in that filter indicates the SW Clump's intrinsic polarization is a factor of about 2 higher than the raw polarization, increasing from 20\% to $\sim$ 40\%.  The SW Clump is a particularly curious feature in the ejecta of VY CMa.  \citet{Humphreys:2007} found it to be moving slowly away from the star with its motion largely in the plane of the sky.  Previously reported AO imaging with LMIRCam has demonstrated that the Clump's 2 $-$ 5 $\micron$ emission must be due largely to diffusely reflected (i.e., optically thick) scattered light \citep{Shenoy:2013}.  

A  silicate grain model fitted to the Clump's 2 $-$ 5 $\micron$ flux by \citet{Shenoy:2013} indicated a dust mass lower limit of 5 $\times$ 10$^{-5}$ M$_{\sun}$, with a temperature constrained to be between 80 $-$ 210 K.  Interestingly, subsequent ALMA millimeter-range (mm) observations of VY CMa reported by \citet{OGorman:2015} did not detect the Clump at 321 \& 658 GHz.  For the lower-limit dust mass computed from the 2 $-$ 5 $\micron$ scattered light, those authors find it would require the SW Clump to have an unusually low temperature ($<$20 K) to be undetected by ALMA.  Alternately, given the 3-$\sigma$ sensitivity limit of their observations, for a temperature between 80 - 210 K they find a dust-mass upper limit of 3 $-$ 7 $\times$ 10$^{-6}$ $M_{\sun}$ within a single 0.229$\arcsec$ $\times$ 0.129$\arcsec$ elliptical beam.   In the infrared the Clump spans an area roughly $\sim1\arcsec$ $\times$ $\sim0.5\arcsec$.  For the area spanned by the beam size of the mm observations, the mass lower limit from scattered light would decrease by a factor of about 3.  Although that brings the infrared and mm mass estimates to within about a factor of 2, some discrepancy between these mass estimates remains.  One possible explanation is the assumed wavelength dependence of the dust emissivity. Typically the emissivity is parameterized with a power-law form where the emissivity ${\varepsilon _\lambda } = (\lambda / \lambda _0)^{ - \beta }$. For the diffuse ISM, $\beta$ takes on the value of 1.6 \citep{Planck-Collaboration:2014}.  \citet{OGorman:2015} use a lower value of $\beta$ = 0.7 based on their own spectral index fits and previous work on mm emission from VY CMa and similar late type stars (e.g. \citet{Knapp:1993}).  \citet{Kaminski:2013} find $\beta$ $\sim$ $-0.5$ for VY CMa's 279 $-$ 355 GHz continuum emission.  They do not see the SW Clump in their continuum map, however they do detect it in their line maps of H$_{2}$S (300.5 GHz) and CS (293.9 GHz).  The absence of mm thermal dust continuum emission from the SW Clump could be due to a value of $\beta$ that is much closer to the ISM value.  The MMT-Pol observations reported here independently reconfirm the previous finding that the Clump is optically thick in the 1 $-$ 3 $\micron$ range (see next subsection), reaffirming the 5 $\times$ 10$^{-5}$ M$_{\sun}$ $minimum$ mass for the SW Clump and underscoring the need for further investigation of this peculiar hypergiant. 

\subsection{Fractional Polarization versus Scattering Optical Depth}
Our polarimetry of IRC +10420 and VY CMa show their nebulae exhibit fairly high intrinsic fractional polarizations in the infrared, with values for the various features ranging from 30\% to 60\% (see Table 2 summary).  High fractional polarization is most easily reconciled with scattering by optically thin dust close to the plane of the sky.  Here we assess whether the polarimetry is consistent with optically thin scattering.  We conservatively define the cut-off for the transition to optical thickness at $\tau^{sc}_{\lambda}$ $\geq$ 0.1.  For reflected light from optically thin dust, the scattering optical depth $\tau^{sc}_{\lambda}$ of each feature may be computed with:
\begin{equation} 
\tau^{sc}_{\lambda} = \frac{ 4\pi \cdot I^{neb}_{\lambda} \cdot \phi^{2}}{F^{\star}_{\lambda} \cdot \sin^{2}(\Theta) \cdot \Phi(\Theta)}
\end{equation}
\citep{Sellgren:1992}.   In this expression $\phi$ is angular radius on the sky, $F^{\star}_{\lambda}$ is the flux of the star at $\lambda$, $\Theta$ is the scattering angle and $\Phi(\Theta)$ is a phase function between 0 and 1 that accounts for the variation with scattering angle $\Theta$ of the intensity of scattered light from a sphere with assumed optical properties.  Since moderately high intrinsic fractional polarizations indicate scattering close to the plane of the sky, we adopt a scattering angle of $\Theta$ = 90$\degr$ (plane of the sky) for simplicity.  Assuming astronomical silicate spheres of size $\bar{a}$ $\lesssim$ 0.3 $\micron$, $\Phi(\Theta = 90\degr)$ $\approx$ 0.4 \citep{Shenoy:2013}.  

For each feature, we use its raw polarized intensity for $I^{neb}_{\lambda}$ because this is a reliable lower-limit on the total scattered light intensity of the feature.  The resulting $\tau^{sc}_{\lambda}$ lower limits discussed here are summarized in Table 2.  For IRC +10420's nebula at 2.2 $\micron$, at the peak of the azimuthal average of the polarized intensity at radius 1.7$\arcsec$ we find $\tau^{sc}_{\lambda}$ $\sim$ 0.4.  For VY CMa, although the star is saturated in the J$'$ images we can estimate its flux $F^{\star}_{J'}$ accurately enough in order to estimate the scattering optical depth of each of the five features discussed in the previous subsection.  We fit a Moffat profile to the unsaturated portion of the star's light profile at a position angle +148$\degr$ E of N, which avoids faint artifacts.  The flux $F^{\star}_{J'}$ obtained from this fit agrees to within $\sim$ 35\% with the star's flux estimated from comparison of the narrow aperture photometry (0.4$\arcsec$ diameter) of VY CMa by \citet{Smith:2001} with the unpolarized standard star HD 224467.  We find $\tau^{sc}_{J'}$ values of 0.7 for the NW Arc, 1.0 for Arc 2, 0.8 for the S Knot, and 0.5 for the S Arc.  For the SW Clump, we find $\tau^{sc}_{J'} >$ 1 and $\tau^{sc}_{3.1}$ = 0.3.  We emphasize that these are \emph{lower limits}, since the polarized intensity of each feature is a lower limit on its total scattered light intensity. 
 
In all cases the computed lower limit to the scattering optical depth is  above a purely optically thin regime.  For grains with very high albedo, it would be expected that multiple scatterings of the light would depolarize it and thus prevent the relatively high intrinsic fractional polarizations seen in both hypergiants' nebulae.  It has been shown in the visual, however, that for reflection from optically thick slabs relatively distant from the illuminating star, the depolarizing effect of multiple scatterings on linear polarization is less than expected at first glance \citep{White:1979}.  The maximum linear polarization of scattered optical light from an optically thick slab is on average within a factor of 0.7 of the polarization for singly scattered light. White demonstrated that dust grains with albedos of $\omega \lesssim 0.4$ could produce fractional polarizations of 40\% in visual light scattered off of an optically thick slab.  For spherical dust grains of radius $a$ $\approx$ 0.1 $\micron$ composed of astronomical silicate \citep{Draine:1984}, the albedo is about 38\% at 2.2 $\micron$.  For spherical dust grains used in modeling dusty oxygen rich, late type stars \citep{Suh:1999}, albedos are well below 40\% at near IR wavelengths.  Our polarization observations are thus consistent with optically thick nebulosity for typical dust.

\begin{deluxetable*}{ccccccc}
\tablecaption{Nebular Features' Intrinsic Polarization and Minimum Scattering Optical Depths}
\tablenum{2}
\tablehead{\colhead{Feature    } & \colhead{PA\tablenotemark{$\dagger$}} & \colhead{distance from $\star$} & \colhead{$\lambda$} & \colhead{$p^{raw}_{\lambda}$} & \colhead{intrinsic $p^{neb}_{\lambda}$} & \colhead{minimum\tablenotemark{$\ddagger$} $\tau^{sc}_{\lambda}$} \\ 
\colhead{} & \colhead{($\degr$ E of N)} & \colhead{($\arcsec$)} & \colhead{($\micron$)} & \colhead{} & \colhead{} & \colhead{} } 
\startdata
& & & & & & \\
\multicolumn{1}{c}{\textbf{IRC +10420}} \\
Azimuthal Avg. & - - & 1.7 & 2.2 & 15\% & 30\% & 0.4 \\
& & & & & & \\
\tableline
& & & & & & \\
\multicolumn{1}{c}{\textbf{VY CMa}} \\
NW Arc  & $-$80 & 3.4 & J$'$ (1.3) & 35\% & 35\% & 0.7 \\
Arc 2 & $-$175 & 4.1 & \textquotesingle\textquotesingle & 45\% & 45\% & 1.0 \\
S Knot & +180 & 1.3 & \textquotesingle\textquotesingle & 40\% & 60\% & 0.8 \\
S Arc & +180 & 2.4 & \textquotesingle\textquotesingle & 40\% & 60\% & 0.5 \\
SW Clump & $-$135 & 1.6 & \textquotesingle\textquotesingle & 35\% & 40\% &  1 \\
\textquotesingle\textquotesingle & $-$135 & 1.6 & 3.1 & 20\% & 40\% & 0.3 
\enddata
\tablenotetext{$\dagger$}{Location of feature with respect to the star.}
\tablenotetext{$\ddagger$}{Scattering optical depths are lower-limits, computed using polarized intensity as a reliable lower-limit on total scattered light intensity (see \S3.3).}
\end{deluxetable*}

\subsection{IRC +10420:  Comparison of K$'$ Polarimetry with 3 $-$ 5 $\micron$ Images}
Here we use the image of IRC +10420's nebula as revealed by the K$'$ (2.2 $\micron$) polarimetry to help interpret the extended emission seen in the LMIRCam 3 $-$ 5 $\micron$ images.  We assess whether the nebula's emission at these longer wavelengths is primarily scattered or thermal light or a combination of both.  The K$'$ polarimetry overlaps with the LMIRCam images starting at a radius $\geq$ 0.5$\arcsec$.  The 3-$\sigma$ levels in the PAH1, PAH2, L$'$ and M images lie at radii of 1.0$\arcsec$, 1.2$\arcsec$, 1.8$\arcsec$, and 1.0$\arcsec$ respectively.   To obtain the broadest wavelength coverage we select an outer radius of 1.0$\arcsec$ for the overlap-of-coverage region (hereafter OCR) that we examine.  The PSF FWHM is $\lesssim$ 0.15$\arcsec$ in all four LMIRCam filters and therefore in the OCR the star's light profile is negligible.  This is in contrast to the MMT-Pol observations at K$'$, where it was necessary to heavily saturate the star image in order to bring up the much fainter nebulosity.  The OCR is depicted with annuli on the LMIRCam images in the bottom row of Figure \ref{LMIRCam_ims} and on the K$'$ image in Figure \ref{IRC_pol_map}(c).    

The nebula's emission from K$'$ through M in the OCR may be a combination of thermal emission and scattered light.  The thermal component is assumed to be unpolarized, while the scattered light can have both unpolarized and polarized components:  
\begin{eqnarray} 
I^{neb}_{\lambda} & = & I^{neb,therm}_{\lambda} + I^{neb,scat} \\
I^{neb}_{\lambda} & = & I^{neb,therm}_{\lambda} + (I^{neb,scat}_{\lambda,U} + I^{neb,scat}_{\lambda,P})
\end{eqnarray}

\begin{figure*} 
\centering
\includegraphics[scale=0.4]{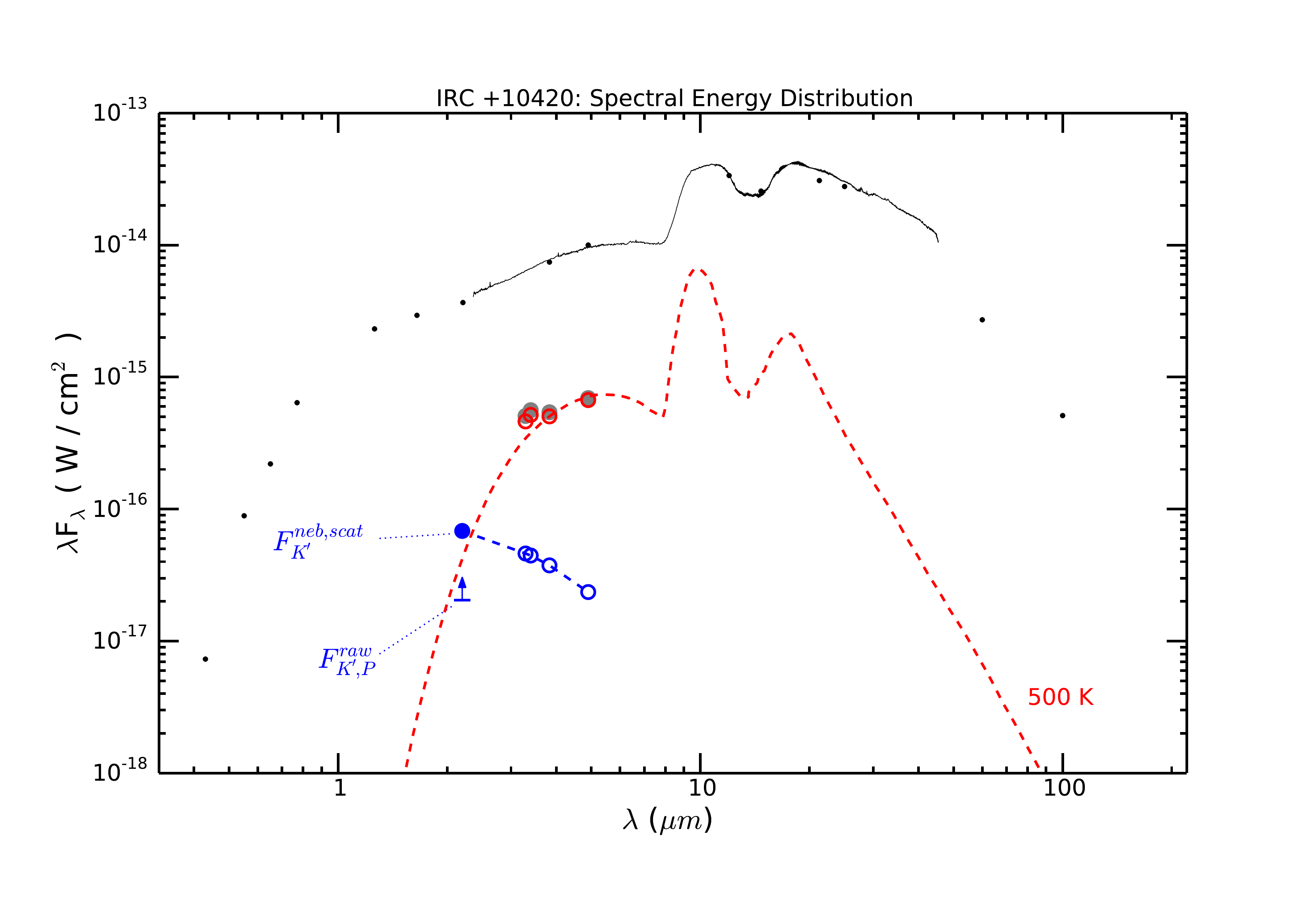}	
\caption{Spectral energy distribution of IRC +10420.  The black dots are point-source photometry compiled from \citet{Jones:1993}, \cite{Oudmaijer:1996} and the IRAS and MSX Point Source Catalogs (with color correction).  The thin black line is the ISO SWS spectrum.   $F^{raw}_{K',P}$ is the polarized intensity at $\lambda$ = K$'$ (2.2 $\micron$) integrated over the overlap-of-coverage region (OCR, the annulus depicted in Figure \ref{IRC_pol_map}(c) and the bottom row of Figure \ref{LMIRCam_ims}).  This is a lower limit on the scattered light flux at K$'$ from this region.  The filled blue circle $F^{neb,scat}_{K'}$ is the estimated total (unpolarized + polarized) scattered light flux for $p^{neb}_{K'}$ $\sim$ 0.3 (\S 4.1).  The blue dashed line scales this total scattered light flux to $\lambda$ = PAH1 (3.29 $\micron$), PAH2 (3.40 $\micron$, L$'$ (3.83 $\micron$) and M (4.9 $\micron$) using Equation (12) in order to predict the nebula's scattered light flux at those wavelengths (open blue circles).  The filled grey circles are the nebula's total 3 $-$ 5 $\micron$ fluxes in those filters, with uncertainty smaller than the symbol size for all except the PAH1 filter.  The open red circles are the excess flux (total $-$ scattered).    The red dashed line is an emissivity-modified $\lambda B_{\lambda}(T_{eq})$ curve for $T_{eq}$ = 500 K. \\} 
\label{IRC_SED}
\end{figure*}

\noindent Subscript $P$ refers to the polarized intensity, and $U$ refers to the unpolarized intensity in the scattered light.   We first consider the minimum and total scattered light flux at K$'$ from the OCR.  As in \S 3.1, the raw polarized intensity at K$'$ (Fig. \ref{IRC_pol_map}(c)) must be scattered light from the nebula: $I^{raw}_{K',P}$ $\longrightarrow$ $I^{neb,scat}_{K',P}$.  This provides a lower limit on $I^{neb,scat}_{K'}$, since the nebula must be less than 100\% polarized.  Assuming that thermal emission at K$'$ is negligible in comparison to the scattered light intensity, then the total scattered light intensity at K$'$ is simply this lower limit divided by the nebula's intrinsic fractional polarization: $I^{neb,scat}_{K'}$ $=$ $I^{raw}_{K',P}$ / $p^{neb}_{K'}$.  Per \S 3.1, $p^{neb}_{K'}$ $\sim$ 0.3, and therefore $I^{neb,scat}_{K'} \sim 3\times I^{neb,scat}_{K',P}$.  The lower-limit and estimated total scattered intensities integrated over the OCR are plotted as fluxes $F^{raw}_{K',P}$ and $F^{neb,scat}_{K'}$ respectively on the star's spectral energy distribution in Figure \ref{IRC_SED}.  

We next consider an upper limit on the contribution of unpolarized thermal emission to the nebular flux at K$'$. Given the polarimetry of VY CMa described in \S3.2 and at visual wavelengths in \citet{Jones:2007} (reproduced in Fig. \ref{VY_pol_map_HST}), the maximum fractional polarization of the scattered light from IRC +10420's nebula at K$'$ we would expect in the OCR is about 60\%: 
\begin{eqnarray} 
p^{neb,scat} & = & \frac{{I_{P}^{neb,scat}}}{{I_{U}^{neb,scat} + I_{P}^{neb,scat}}} \le 0.6 \\
		    & = & \frac{{I_{P}^{neb,scat}}}{{I^{neb,scat}}} \le 0.6
\end{eqnarray}
The nebula's intrinsic fractional polarization of 30\% (\S3.1) is determined by the nebula's total scattered plus thermal emission:
\begin{equation} 
p^{neb} = \frac{{I_{P}^{neb,scat}}}{{I^{neb,scat} + I^{neb,therm}}} = 0.3
\end{equation} 
Comparing the maximum and intrinsic polarizations, we can write:  
\begin{equation} 
I^{neb,therm} \le I^{neb,scat}
\end{equation}
Thus, the maximum contribution of unpolarized thermal emission to the observed flux from the OCR in the $K'$ filter would be roughly equal to the scattered light contribution. 

The intensity of scattered light at a given wavelength is directly proportional to the illuminating star flux, which we designate as $F^{\star}_{\lambda}$, and the grains' albedo $\omega_{\lambda}$.  We scale $I^{neb,scat}_{K'}$ from K$'$ to the four longer wavelengths $\lambda$ = PAH1, PAH2, L$'$ and M using ratios for the illuminating flux and grain albedo at the two wavelengths. We are assuming the OCR is outside of the region producing the bulk of the emission that makes up the SED of IRC+10420. In this way we can take the observed SED as the illuminating flux $F^{\star}_{\lambda}$ and express the following relationship:  

\begin{equation} 
I^{neb,scat}_{\lambda} = I^{neb,scat}_{K} \left( \frac{F^{\star}_{\lambda}}{F^{\star}_{K}} \right) \left( \frac{\omega_{\lambda}}{\omega_{K}} \right) 
\end{equation}

\noindent Note that the effects of interstellar reddening on the star's SED and the nebula are the same, so the extrapolation from K$'$ to the longer wavelengths can be done directly with the reddened (observed) SED.  The ratio of the observed fluxes is $F^{\star}_{\lambda}/F^{\star}_{K'}\sim  2$ for all four LMIRCam filters.

Assuming silicate grains of radius $a$ $\ll$ $\lambda$ (e.g., $a$ $\lesssim$ 0.1 $\micron$) the ratio of the albedos $\omega_{\lambda}/\omega_{K'}$ for the four filters are 0.39, 0.36, 0.26, and 0.13. These albedo ratios are typical ratios for astronomical silicates for the interstellar medium \citep{Draine:1984} and oxygen rich mass-loss winds \citep{Suh:1999}.  With these values in Equation (12), we predict the 3 $-$ 5 $\micron$ scattered light fluxes depicted with open blue circles in Figure \ref{IRC_SED}. On the SED in Figure \ref{IRC_SED}, the nebula's integrated 3 $-$ 5 $\micron$ fluxes for the OCR from 0.5$\arcsec$ to 1.0 $\arcsec$ are plotted as grey circles.  The open red circles are the excess flux in each of these 3 filters (total $-$ predicted scattered). As can be seen in Figure \ref{IRC_SED}, the total nebular emission at 3 $-$ 5 $\micron$ is about a factor of 10 brighter than scattered light alone can account for.  It is most likely the nebula's emission at $\sim$ 3 $\micron$ and beyond is primarily thermal in origin. However, we have just shown that at most half of the flux at $2.2~\micron$ can be thermal, and this puts strong constraints on any explanation for the surprisingly large 3 $-$ 5 $\micron$ fluxes. 

On the SED in Figure \ref{IRC_SED} we plot the spectrum of a 500 K emissivity-modified blackbody which passes through the L$'$ excess flux. This choice of temperature was chosen to prevent too much thermal flux at $2.2~\micron$ (a higher temperature) and too much flux in the mid-IR (a lower temperature) which would compete with or exceed the observed mid-IR flux from the entire system. A lower temperature such as 400 K, for example, would require fully half of the entire $10~\micron$ silicate feature to arise from the OCR alone, which seems unlikely. The blackbody equilibrium temperature is 120 K, assuming $L^{\star}_{bol}$ = 5 $\times$ 10$^{5}$ $L_{\sun}$ and $D$ = 5 kpc.  Factoring in wavelength-dependent emissivity $Q^{abs}_{\lambda}$  to account for less efficient emission in the infrared, the dust equilibrium temperature is $T_{eq}$ = $\left( \langle Q_{UV,vis} \rangle / \langle Q_{IR} \rangle \right)^{1/4}T_{bb}$, where $\langle Q_{UV,vis} \rangle$ and $\langle Q_{IR} \rangle$ are the Planck-averaged emissivities for the wavelength ranges where the grains absorb and emit, respectively. For the typical silicate grains we have been using \citep{Draine:1984, Suh:1999} this factor raises the equilibrium temperature to $\sim$ 210 K.  Although warmer than blackbody equilibrium, this is still too cool to be compatible with the observed mid-IR flux. 

The $\sim 500$ K color temperature we require is 2 to 3 times higher than the expected equilibrium temperature. One possibility is emission from transiently heated small dust grains, which can temporarily reach much higher equilibrium temperatures from absorption of a single stellar photon with an energy comparable to the grain's heat capacity. These grains are usually associated with reflection nebulae illuminated by a source of near-UV photons, e.g., \citet{Sellgren:1983}.   However, the spectrum of emission from these very small grains corresponds to color temperatures much higher than the 500 K we require to explain the emission from the OCR.  For example, \citet{Draine:2007} compute SEDs for $a$ = 5 \AA ~silicate grains exposed to the interstellar radiation field (roughly equivalent to an F0 star, similar to the IRC+10420).  They find such a tiny grain's spectrum to be relatively flat from $2.2 - 5~\micron$ (see their Figure 10), which is incompatible with our observations. We do note that our 500 K color-temperature is close to with the {\bf gas} temperature at this radius ($r \sim 0.75\arcsec = 5.6 \times 10^{16}$ cm) modeled by \citet{Castro-Carrizo:2007} in their study of CO emission around IRC +10420. These authors modeled the CO emission as arising from two spherical shells, with the inner shell spanning from 0.3$\arcsec$ to 1.7$\arcsec$ in radius, very similar to our OCR.  At the 0.75$\arcsec$ mean radius of the OCR we examine, their best-fit model predicts a {\bf gas} temperature of $T_{eq}$ = 460 K. While our polarimetry observations clearly require a nebular material lying largely in the plane of the sky as opposed to a shell, the similarity in their gas temperature and the dust color temperature we derive indicates that higher than equilibrium temperatures are physically real. 

\section{CONCLUSIONS}
1. We present images of the nebulosity surrounding the hypergiant VY CMa in polarized light with a resolution of $0.2\arcsec$ at 1.3, and $3.1~\micron$. Many nebular features are seen in polarized intensity at $1.3~\micron$ and one feature, the SW Clump, is very prominent at $3.1~\micron$ as well.

2. Extended emission around VY CMa shows both high fractional polarization and high scattering optical depths. The surface brightness is consistent with scattered light alone. The high fractional polarization is consistent with grain albedos considerably less than 1. The required albedos are easily compatible with the typical silicate grains used to model dusty stars.

3.  The polarimetry of VY CMa independently confirms that its Southwest (SW) Clump is optically thick from 1.3 $-$ 3.1 $\micron$.  This affirms the minimum mass of the SW Clump computed by \citet{Shenoy:2013} from its scattered light in the infrared.  Absence of thermal dust emission at mm wavelengths from the SW Clump presents a puzzle.

4. We present images of the nebulosity surrounding IRC+10420 in polarized light at $2.2~\micron$ with a resolution of $0.2\arcsec$ and total flux at 3.3, 3.4, 3.8, and $4.9~\micron$ with a resolution of $0.1\arcsec$ $-$ 0.15$\arcsec$. 

5. The polarized intensity image of IRC+10420 at $2.2~\micron$ shows a relatively uniform nebula largely in the plane of the sky extending from $0.5\arcsec-2.5\arcsec$ radius (2500 $-$ 12500 AU for D = 5 kpc). The surface brightness of this low-latitude ejecta is compatible with optically thick scattering and, similar to VY CMa, grain albedos compatible with typical astronomical silicates. 

6. In the $3-5~\micron$ band, images of IRC+10420 show a strong extended component that overlaps with the nebula seen in polarized intensity at $2.2~\micron$. The flux from this extended component is an order of magnitude brighter than can be explained by simple extrapolation of the scattered light seen at $2.2~\micron$. We hypothesize grains warmed to a temperature higher than the expected grain equilibrium temperature, but consistent with the local gas temperature in this region.

\acknowledgements
The authors thank the referee Dr. Geoffrey Clayton for his thoughtful comments which have helped us to improve this manuscript.  We also thank Dr. Roberta M. Humphreys for many illuminating discussions and her insights on the nature and history of these fascinating targets.  We gratefully acknowledge the steady support of the staffs of the MMT and LBT Observatories in making these observations possible.  MMT-Pol is funded by the National Science Foundation under grant NSF AST-0705030.    LMIRCam is funded by the National Science Foundation under grant NSF AST-07049992.

\bibliographystyle{apj}
\bibliography{ms.bbl}

\end{document}